# Physical properties revealed by transport measurements on superconducting $Nd_{0.8}Sr_{0.2}NiO_2$ thin films

Ying Xiang[a], Qing Li[a], Yueying Li[b], Huan Yang[a,*], Yuefeng Nie[b], Hai-Hu Wen[a,*]

[a] *National Laboratory of Solid State Microstructures and Department of Physics, Center for Superconducting Physics and Materials, Collaborative Innovation Center for Advanced Microstructures, Nanjing University, Nanjing 210093, China*

[b] *National Laboratory of Solid State Microstructures, Jiangsu Key Laboratory of Artificial Functional Materials, College of Engineering and Applied Sciences,*

*Collaborative Innovation Center of Advanced Microstructures, Nanjing University, Nanjing 210093, China*

\* Corresponding authors.

*E-mail addresses*: huanyang@nju.edu.cn (H. Yang), hhwen@nju.edu.cn (H. H. Wen)

**Abstract**

The newly found superconductivity in infinite-layer nickelate superconducting films has attracted much attention, because their crystalline and electronic structures are similar to high-$T_c$ cuprate superconductors. The upper critical field can provide much information on superconductivity, but detailed experimental data are still lacking in these films. Here we present temperature and angle dependence of resistivity measured under different magnetic fields ($H$) in $Nd_{0.8}Sr_{0.2}NiO_2$ thin films. The onset superconducting transition occurs at about 16.2 K at 0 T. Temperature dependent upper critical fields determined by using a criterion very close to the onset transition show a clear negative curvature near the critical transition temperature, which is explained as the consequence of the paramagnetically limited effect on superconductivity. The temperature dependent anisotropy of the upper critical field is obtained from resistivity data, which yields a value decreasing from 3 to 1.2 with lowering temperature. This can be explained by a variable contribution from the orbital limit effect on upper critical field. The angle dependent resistivity at a fixed temperature and different magnetic fields cannot be scaled to one curve, which deviates from the prediction of the anisotropic Ginzburg-Landau theory. However, at low temperatures, the increased resistivity by magnetic field can be scaled by the parameter $H^{\beta}|cos\theta|$ ($1 < \beta < 6$) with $\theta$ the angle enclosed between $c$-axis and the applied magnetic field. As the first detailed study on the upper critical field of the nickelate thin films, our results clearly indicate a small anisotropy and paramagnetically limited effect of superconductivity in nickelate superconductors.

**Keywords:** Nickelate superconductor; Transport measurements; Upper critical field; Paramagnetically limited superconductor

## 1. Introduction

Superconductivity was successfully detected in Nd$_{1-x}$Sr$_x$NiO$_2$ and Pr$_{1-x}$Sr$_x$NiO$_2$ thin films recently [1,2], and the observation is very important because the nickelates probably share similar electronic structures with that of high-$T_c$ cuprate superconductors. Until now, superconductivity has only been observed in thin films grown on SrTiO$_3$ substrates [1-5], but has not been observed in Nd$_{1-x}$Sr$_x$NiO$_2$ bulk samples [6,7]. The superconducting transition occurs in a narrow doping range of Sr, namely 0.125 < $x$ < 0.25 in Nd$_{1-x}$Sr$_x$NiO$_2$ films. Meanwhile, the underdoped and overdoped samples seem to show a weak insulating property [4,5]. The multiband nature of the material is proved experimentally by the temperature dependent Hall coefficient [1,4] and electron energy loss spectroscopy results [8]; it is also supported by several theoretical calculations of electronic structures [9-27]. In the parent compound of NdNiO$_2$, three sets of three dimensional (3D) Fermi pockets are predicted by the theoretical calculations [13-16], i.e., a big α Fermi surface contributed by the Ni-3$d_{x^2-y^2}$ orbital and the two small electron pockets (β and γ) contributed by a mixed orbital contribution from Ni- $3d_{3z^2-r^2}$ and Nd- $5d_{3z^2-r^2}$ / $5d_{xy}$ orbitals due to a considerable hybridization effect [23]. The α pocket crosses the whole Brillouin zone in $k_z$ direction in NdNiO$_2$, and it behaves [13-15] as a hole and an electron pocket at the cut of $k_z$ = 0 and $k_z$ = π, respectively. Besides, the low-energy Nd 5d states have a self-doping effect to Ni-3$d_{x^2-y^2}$ orbital [16-18], which may form a Kondo spin singlet state and make the electronic structure of Ni ion different from the Cu$^{2+}$ in cuprates. With 20% Sr doping in NdNiO$_2$, theoretical calculations tell that the self-doping effect may disappear making the material more cuprate-like [16,17]; the shapes of Fermi pockets change when compared with the undoped sample [25-27]. Near $k_z$ = 0, the hole-like part of the α pocket becomes a bit larger in the doped sample; while near $k_z$ = π, the electron-like part of the α pocket becomes a bit smaller [13,25-27]. The other two electron pockets shrink or even disappear with doping [25-27]. Thus, there are still two [25-27] or three [13] sets of 3D Fermi surfaces mainly contributed by Ni 3$d$ and Nd 5$d$ electrons in Nd$_{0.8}$Sr$_{0.2}$NiO$_2$. In this point of view, it is very interesting to investigate the superconducting anisotropy and the critical behavior in the infinite-layer nickelate superconductors with multiband effect.

As a newly found superconducting system, the possible origin of the superconductivity is discussed in several related works [9,12,15,28-31], and the possible gap symmetry is discussed based on theoretical calculations [13,14,32,33]. A recent scanning tunneling microcopy (STM) work shows that there are two types of superconducting gaps in the system [34], i.e., a *d*-wave gap with a gap maximum of about 3.9 meV and a slightly anisotropic *s*-wave gap with a gap maximum of about 2.35 meV. The presence of a *d*-wave gap further strengthens the believe of similarity between nickelate superconductors and cuprates. In most superconductors [35], the superconducting gap ($\Delta$) and the pairing strength may be linked to the upper critical field $\mu_0 H_{c2}$ via the Pippard relation $\xi = \hbar v_F/\pi\Delta$ and $\mu_0 H_{c2} = \Phi_0/2\pi\xi^2$. Here, $\xi$ is the coherence length, $v_F$ is the Fermi velocity, and $\Phi_0$ is the flux quantum. However, in a few superconductors, Cooper pairs break mainly due to the Zeeman-split effect [36], while the upper critical field is dominated by Pauli paramagnetic limit $\mu_0 H_P^{pair} = \sqrt{2}\Delta/g\mu_B$. Here $\mu_B$ is the Bohr magneton, and *g* is the Landé factor. Therefore, it is worthy measuring the upper critical field to obtain the information of the pairing strength for this new superconducting system.

Herein, we report our experimental results of the temperature and angle dependent resistivity measured at different magnetic fields in superconducting $Nd_{0.8}Sr_{0.2}NiO_2$ thin films. We observe a negative curvature on $\mu_0 H_{c2}$-*T* curves near $T_c$ when *H* || *ab*-plane or *H* || *c*-axis *c*-axis. It is further found that the angle-dependent resistivity cannot be scaled by the anisotropic Ginzburg-Landau (GL) theory. These results suggest exotic properties of superconductivity in $Nd_{0.8}Sr_{0.2}NiO_2$ thin films.

**2. Materials and methods**

The $Nd_{1-x}Sr_xNiO_3$ thin films were grown on $SrTiO_3$ substrates by using the reactive molecular beam epitaxy technique with a nominal composition of *x* = 0.2. The thickness of the film is about 6 nm. We then use the soft-chemistry topotactic reduction method [1,5] to remove the apical oxygen and obtain the superconducting $Nd_{1-x}Sr_xNiO_2$ phase. In the beginning of the topotactic hydrogen procedure, a precursor $Nd_{0.8}Sr_{0.2}NiO_3$ thin film was placed in a quartz tube together with a pellet of $CaH_2$ weighted about 0.5 g. The tube was evacuated and sealed, then it was annealed at 340°C for 100 min. There is no direct contact between the samples and $CaH_2$ during the treatment process. The structure of the resultant $Nd_{0.8}Sr_{0.2}NiO_2$ film is characterized by the appearance of the (001) and (002) peaks in x-ray diffraction data measured by a Bruker D8 Discover diffractometer. The *c*-axis of the film is perpendicular to the substrates. The resistivity was measured by using a standard four-electrode method in a physical property measurement system (PPMS, Quantum Design) with magnetic fields up to 9 T.

**3. Results**

*3.1. Upper critical field*

Figure 1 shows the temperature dependent resistivity measured at different magnetic fields when the field is parallel or perpendicular to the *c*-axis of the film. One can see that the normal-state $\rho$-$T$ curve shows an almost linear behavior with a positive slope when $T > 20$ K. The normal state resistivity $\rho(T = 20\ \text{K}) = 0.38\ \text{m}\Omega\cdot\text{cm}$, and the corresponding residual resistance ratio $\rho(T = 300\ \text{K})/\rho(T = 20\ \text{K}) = 2.8$ determined from the wide-temperature-range $\rho$-$T$ curve (not shown here). Resistivity curves in Fig. 1 show a very weak magnetoresistance in the presence of a magnetic field of up to 9 T. The magnetoresistance value is only +0.16% at $T = 20$ K and $\mu_0 H = 5$ T, and this value is similar to the one from a previous report [4]. A negative Hall coefficient $R_H = -(2.7\pm0.3)\times10^{-3}\ \text{cm}^3/\text{C}$ is obtained from the transverse resistance measurement by using a standard six-electrode method at $T = 20$ K and with the maximum field of 5 T. The negative Hall coefficient is different from positive values reported previously [4,5]. The different signs of Hall coefficient may be due to slightly different oxygen contents in different films, and the sign can be easily changed in this material with almost balanced charge densities of holes and electrons. Derived from the temperature dependent resistivity measured at 0 T and shown in Fig. 1, the onset superconducting transition temperature $T_c^{\text{onset}}$ at 0 T is about 16.2 K determined by the criterion of $95\%\rho_n(T)$, and $\rho_n(T)$ is the linear extrapolation of the normal-state resistivity. The zero-resistance transition temperature $T_{c0}$ is about 9.3 K determined with the criterion of $1\%\rho_n(T)$. Here, the transition width is about 7 K determined from the $\rho$-$T$ curve measured at 0 T in the film. There are two possibilities for such a broad transition in zero field: either a lot of disorders or a very non-mean-field transition which would point towards an unconventional superconductivity. Since films are usually much more disordered than single crystals, the inhomogeneity or disorder effect may be the primary reasons for such a broad transition in the studied films. And the statement is supported by the large normal-state residual resistivity of the film.

In Fig. 1, one can see that the transition temperature decreases and the transition width widens when the magnetic field is applied in two perpendicular directions. In order to have a quantitative analysis on the field dependent critical temperatures, we try to obtain the values of zero-resistance transition fields $\mu_0 H_0$ and $\mu_0 H_{c2}$ from the $\rho$-$T$ curves by using different criterions. Figure 2a,b show temperature dependent characteristic fields. In our point of view, the zero-resistance transition fields $\mu_0 H_0$ is determined by the phase with the lowest $T_c$ and is also possibly affected by some weak-link behavior between different superconducting regions. Therefore, it is difficult to obtain reliable information of the irreversibility field $\mu_0 H_{\text{irr}}$ from the data by using a small resistivity criterion. However, the upper critical field obtained by using a high value of resistivity criterion should be dominated by the superconducting phase with the highest $T_c$, and such phase should hold the strongest upper critical field. Hence, the feature of the upper critical field should be intrinsic.

Obviously, the slopes of $\mu_0 H_{c2}$-$T$ curves are huge near $T_c$. In addition, one can see in Fig. 2a,b that the slope $\mu_0 dH_{c2}/dT$ decreases with the increase of temperature in both cases of $H \| c$-axis and $H \| ab$-plane, showing a clear negative curvature of $H_{c2}(T)$. It is known that in a 2D superconducting system, $\mu_0 H_{c2} \propto (1-T/T_c)^{1/2}$ may appear when the magnetic field is parallel to the film plane [37], which certainly leads to a negative curvature. However, the negative curvature does not show up near $T_c$ when the field is perpendicular to the film [37]. From our data, the negative curvature appears on the $\mu_0 H_{c2}$-$T$ curves when the magnetic field is along or perpendicular to the film plane, which excludes the possibility of the 2D superconductivity. We also try to treat the experimental data from previous reports [1,5] when the magnetic field is perpendicular to the films ($H \| c$-axis), and similar negative curvatures are obtained near $T_c$ on $\mu_0 H_{c2}$-$T$ curves by using the criterion of 95%$\rho_n(T)$ (data and treatment not shown here). In these previous works [1,5], thicknesses of the films are 11 nm and 35 nm, respectively, which means that the negative curvature seems to be a common feature in the material even when the magnetic field is along $c$-axis of the sample.

Since the normal-state residual resistivity is large in the film, we try to use the Werthamer, Helfand, and Hohenberg (WHH) theory [36] in dirty limit for superconductor with a single $s$-wave gap to fit our experimental data. Then the upper critical field can be derived based on the WHH theory [36] via

$$\ln\frac{1}{t} = \sum_{v=-\infty}^{\infty}\left\{\frac{1}{|2v+1|} - \left[|2v+1| + \frac{\bar{h}}{t} + \frac{(\alpha_M \bar{h}/t)^2}{|2v+1| + (\bar{h}+\lambda_{SO})/t}\right]^{-1}\right\} \quad (1)$$

Here $t = T/T_c$; $\bar{h} = 4\mu_0 H_{c2}(T)/(\pi^2 H' T_c)$ with $H' = \mu_0|dH_{c2}/dT|_{T_c}$; $\lambda_{SO}$ is the parameters presenting the strengths of the spin-orbit interaction; $\alpha_M = \sqrt{2} H_{orb}/H_P$ is the ratio of the orbital limit $H_{orb}$ over the paramagnetic limit $H_P$, describing the contribution ratio between the pairing-breaking Zeeman energy and the orbital pair-breaking energy [38,39]. We try to calculate $H'$ by using the $\mu_0 H_{c2}(T)$ data obtained at fields of 0 and 1 T, and obtained values are given in Table 1. By assuming the absence of spin-orbital interaction ($\lambda_{SO} = 0$) and in the situation of $\alpha_M = 0$, we can plot the theoretical curves of $\mu_0 H_{c2}(T)$ purely contributed by orbital. These curves are shown as dashed lines in Fig. 2a,b. One can see almost linearly temperature dependent upper critical field in the displayed range; the approximately linear behavior can be observed in the fitting curve when the temperature is from $0.7T_c$ to $T_c$ from previous theoretical calculations [36,40]. However, the linear part of the experimentally obtained $\mu_0 H_{c2}(T)$ data is really very narrow, which is obviously deviated from theoretical curves when the upper critical field is only contributed by the orbital. The orbital limited upper critical field can be calculated by the empirical formula $\mu_0 H_{orb}(0) = -0.69 T_c H'$ in dirty limit [40] which is equivalent to use Eq. (1) to obtain the upper critical field in zero temperature limit in the condition of $\alpha_M = 0$. The calculated values of $\mu_0 H_{orb}(0)$ are listed in Table 1. Although the experimental data obviously deviated from the theoretical model when $\alpha_M = 0$, one can still obtained the orbital limit $\mu_0 H_{orb}(0)$ from the slope $H'$ at temperatures very close to $T_c$ based on the previous theoretical work [36]. The anisotropic ratio in Table 1 is defined as $\Gamma = H_{c2}^{H||ab}/H_{c2}^{H||c}$ with $H_{c2}^{H||ab}$ and $H_{c2}^{H||c}$ representing the upper critical fields along $ab$-plane and $c$-axis, respectively. These anisotropic ratios are not very large, and the value is near the upper limit of the measured $\Gamma$ (Fig. 2c).

Since the WHH theory with $\alpha_M = 0$ cannot agree with the experimental data of $\mu_0 H_{c2}(T)$, we then try to fit the data with the only fitting parameter of $\alpha_M$ by assuming $\lambda_{SO} = 0$. Here $H'$ has already been obtained as mentioned above. One can see that the solid lines in Fig. 2a,b fit the experimental data well, and the obtained $\alpha_M$ in Table 1 is very large in the material. From the values of $\mu_0 H_{c2}^{orb}(0)$ and $\alpha_M$, we can also get the values of $\mu_0 H_P(0)$ and show the values in Table 1. The upper critical field in the zero-temperature limit [36] $H_{c2}(0) = H_{orb}(0)$ when $\alpha_M \to 0$, while $H_{c2}(0) = H_P(0)$ when $\alpha_M \to \infty$. The relatively small value of $\mu_0 H_P$ may confirm that paramagnetic pair breaking dominates the upper critical field in the sample. The anomalous negative curvature on the $\mu_0 H_{c2}$-$T$ curve near $T_c$ as well as the large value of $\alpha_M$ may suggest a dominant paramagnetic pair breaking effect in the present thin films, which is similar to the situation in other paramagnetically limited superconductors [41-57]. In addition, the obtained value of $\alpha_M$ is even much larger than those obtained in most heavy-fermion and organic superconductors [43-49,54]. A very large value of $\alpha_M$ indicates a dramatic mismatch between $H_{orb}$ and $H_P$. It also indicates the possible existence of the Fulde-Ferrell-Larkin-Ovchinnikov (FFLO) state [58,59] in the high-magnetic-field region at low temperatures when the magnetic field gives a more influential role in breaking the spin-pairing. Since the FFLO state is fragile in the presence of disorder [54], the existence of this state in the $Nd_{0.8}Sr_{0.2}NiO_2$ thin film requires further high-magnetic-field experiments.

*3.2. Angular dependent resistivity*

The anisotropic upper critical field can usually be observed in angle dependent resistivity. In Fig. 3, we show the angle dependence of resistivity measured at different temperatures and magnetic fields. Being different from the data measured in other systems, the resistivity dip near θ = 90° (H ∥ ab-plane) is very sharp. Based on the anisotropic GL theory [35], angle dependence of the orbital limiting upper critical field can be expressed as

$$H_{c2}(\theta) = \frac{H_{c2}^{H||c}}{\sqrt{\cos^2\theta + \Gamma^{-2}\sin^2\theta}} \quad (2)$$

Then the angle resolved resistivity can be scaled [60] with the effective field $\widetilde{H} = H\sqrt{\cos^2\theta + \Gamma^{-2}\sin^2\theta}$ by adjusting the anisotropic ratio Γ. The theory successfully works in the iron-based [61,62] and BiS$_2$-based [63] superconductors of very different anisotropy ratios. In the scaling procedure, the scaled curves should coincide to the field dependent resistivity when H//c-axis (θ = 0) because $\widetilde{H} = H$ in this direction [64]. It should be noted that the anisotropic GL theory is derived from the anisotropic vortex pinning strength [60], but the anisotropic orbital limiting upper critical field is used in the scaling process. As mentioned above, the upper critical field may be dominated by the Pauli paramagnetic limit in the Nd$_{0.8}$Sr$_{0.2}$NiO$_2$ thin film, and here we want to check whether the anisotropic GL theory can describe the angle dependence of resistivity in the vortex pinning picture. However, the scaling is not successful with any values of Γ. One set of examples of failed fittings is shown in Fig. 4 with Γ = 2, and it is impossible to make all the curves scale together and the scaling curves deviate from ρ-H curve with θ = 0. In this point of view, our results obviously deviate from the standard anisotropic GL scaling theory.

The failed scaling behavior has been found in repeated experiments measured in the other two films. The first possibility for the failed scaling may be inhomogeneous superconducting phases in the film. The second possibility may be that there are 2 or 3 sets of Fermi surfaces with different anisotropies [13,25-27] in the material, while the scaling model is assumed to be applicable in the single band situation. The third possibility may be that the film is very thin, which would affect the parallel component of $\mu_0H_{c2}$ and thus contribute to the angular dependence of $\mu_0H_{c2}$, particularly at temperatures near $T_c$ where the coherence length diverges. However, it should be noted that Eq. (2) is used to describe the anisotropic behavior of the orbital limiting upper critical field $\mu_0H_{c2}^{orb}(\theta)$; therefore, it is natural that the scaling does not work because the upper critical field is dominated by the paramagnetic limit $\mu_0H_P$ here.

It is difficult to obtain a simple function of $H_{c2}(\theta)$ in the paramagnetically limited system

[41], so we try to scale the measured data in other ways. Here we use a new scaling parameter $(\mu_0 H)^\beta |\cos\theta|$ to scale the increased resistivity by the magnetic field $\Delta\rho(H, \theta) = \rho(H,\theta) - \rho(H,\theta = \pi/2)$. The scaling results are presented in Fig. 5. The new scaling law seems to work well for the data taken at 6 and 8 K, although $\beta$ decreases with increase of temperature. While the scaling becomes worse at 10 K and fails again for data measured at higher temperatures. The applicability of the new scaling law confirms the failure of the scaling by the anisotropic GL theory, in the latter a zero resistance should appear and ramps gradually due to the dissipation of vortex motion near the angle $\theta = \pi/2$. It seems that the $c$-axis component of the external magnetic field is more influential to enhance resistivity than that expected by the anisotropic GL theory. The new proposal for the scaling in Fig. 5 is just a surprising empirical discovery, which needs the theoretical support and further studies with better quality samples, such as single crystals.

## 4. Discussion

In the superconducting $Nd_{0.8}Sr_{0.2}NiO_2$ films, we observe a negative curvature in $\mu_0 H_{c2}$-$T$ curves near $T_c$ when the magnetic field is parallel to the $ab$-plane or the $c$-axis, and the angular dependent resistivity measured at different fields cannot be scaled according to the anisotropic GL theory. Since the negative curvature appears in $\mu_0 H_{c2}$-$T$ curves when the magnetic field is along the $c$-axis from our data and previous reported data [1,5] measured in films with different thicknesses from 6 to 35 nm, this contradicts the expectation for a 2D superconducting sample (as argued above), thus the possibility from the 2D superconductivity may be excluded. There are two possible reasons to interpret these experimental results. One is the inhomogeneity of the film. In this case, the transition width is large, and the film may be composed by different superconducting phases with distinct $T_c$ values. However, we argue that the upper critical fields determined by the method adopted here for $Nd_{0.8}Sr_{0.2}NiO_2$ films still reflect the intrinsic properties of the sample. This statement is based on following arguments: (1) Given the inhomogeneity of the film, the upper critical field obtained by using a high value of resistivity criterion (e.g. 98%$\rho_n$) should be dominated by the superconducting phase with the highest $T_c$, and this phase should hold the strongest upper critical field. (2) The negative curvature in $\mu_0 H_{c2}$-$T$ curves near $T_c$ can also be observed in other two sets of reported data from different groups [1,5], thus we conclude that such feature is related to the intrinsic superconducting mechanism of the material. The second possibility of our observations may be that both upper critical field and the field induced resistivity are determined by the Zeeman pair breaking effect. In this situation, the upper critical field in paramagnetic limit can be estimated from the binding energy of Cooper pairs, i.e., $\mu_0 H_P^{pair} = \sqrt{2}\Delta/g\mu_B$. Based on our recent STM work [34], two superconducting gaps may open in different Fermi surfaces, and the gap functions read $\Delta_d = 3.9\cos2\phi$ meV (a $d$-wave gap) and $\Delta_s = 2.35(0.15\cos 4\phi + 0.85)$ meV (a slightly anisotropic $s$-wave gap), respectively. Averaged gap values determined by $\bar{\Delta}^2 = \frac{1}{2\pi}\int_0^{2\pi}\Delta^2(\phi)d\phi$ are $\bar{\Delta}_d = 2.76$ meV and $\bar{\Delta}_s = 2.01$ meV. By assuming the Landé factor $g = 2$ for free electrons, the corresponding $\mu_0 H_{P\,d}^{pair} = 21$ T and $\mu_0 H_{P\,s}^{pair} = 15$ T, respectively. These estimated values of paramagnetic limits are very close to the fitting results given in Table 1. It should be noted that the estimation based on the single gap model is very crude for this complex material, but the estimation results justify our analysis here.

A related observation is the incredibly large value of $\alpha_M$ obtained from the fittings. The reason of the large $\alpha_M$ in heavy-fermion superconductors is due to the large effective mass. Based on the single-band model in the clean limit of Bardeen-Cooper-Schrieffer (BCS) theory, the Maki parameter $\alpha_M \propto m\Delta/E_F$ when $H \parallel c$-axis [38,39]. A very large value of $\alpha_M$ corresponds to a very large effective mass $m$ and/or a very small Fermi energy $E_F$. However, both parameters are lacking for superconducting nickelate films at this moment. Theoretically, it was suggested that [18] the Nd-5$d$ conduction electrons may couple to localized Ni-3$d_{x^2-y^2}$ electrons to form Kondo spin singlets, which leads to an enhanced effective mass at low temperatures. The discussions above are based on the single-gap model, while the true situation is that it is a multiband superconductor, which makes the analysis more complicated. For example, there are many more fitting parameters to fit the $\mu_0H_{c2}$-$T$ data in the multi-gap situation than in the single-gap one [57,65]. A good fitting by multi-gap model requires more data of $\mu_0H_{c2}$-$T$ measured at much higher fields and low temperatures. Therefore, it is highly desired to measure the upper critical field directly at low temperatures and high magnetic fieldsin future works.

Another observation from our experiments is that the anisotropy ratio is not large for both $\mu_0H_P$ and $\mu_0H_{orb}$ in these films. The obtained $\Gamma$ is in the range of 1.2 to 3.0 as derived from the experimental data in Fig. 2c, while $\Gamma_{orb}(0)$ is about 2.3-2.97 from fittings. The anisotropy ratio is comparable to the ones in iron-based superconductors [66] but much smaller than the ones in cuprates, like Bi-2212 system [35] and BiS$_2$-based [63] superconductors. It should be noted that the Nd$_{1-x}$Sr$_x$NiO$_2$ material is the infinite-layer phase; therefore, it is understandable for a small anisotropy value. In addition, many theoretical calculations [13,25-27] illustrate that there is a big α Fermi surface with a strong dispersion along $k_z$-axis in $Nd_{0.8}Sr_{0.2}NiO_2$, meanwhile there is still one [25-27] or two [13] small 3D electron Fermi pockets, all these will lead to a low anisotropy. Based on the picture of a paramagnetically dominated superconductivity, $\Gamma_P(0)$ is only about 1.25-1.29 from the fitting, which suggests a small anisotropy of Pauli susceptibilities or $g$ factors along the two perpendicular axes [41]. The increase of the anisotropy in Fig. 2c with temperature can be explained as the increased contribution from the orbital limit when temperature approaches to $T_c$. The orbital limiting effect dominates the upper critical field near $T_c$ and generally gives a relatively larger anisotropy. On the contrary, in the low temperature region, the Pauli limit should give a relatively smaller anisotropy, play an increasingly important role, and finally become decisive in determining the upper critical field. That explains the strong temperature dependence of the anisotropy of measured upper critical fields.

**5. Conclusion**

In summary, we have conducted resistive transport measurements under magnetic fields in superconducting thin films of $Nd_{0.8}Sr_{0.2}NiO_2$ with the onset superconducting transition temperature of about 16.2 K. The anisotropy of the measured upper critical field is small, locating in the range of 1.2 to 3 near the transition temperature. We observe a negative curvature of the $\mu_0H_{c2}$-$T$ curve near $T_c$, which is explained as the possible consequence of the paramagnetic limited superconductivity. The angle dependence of resistivity at a fixed temperature and different magnetic fields cannot be scaled by using the anisotropic Ginzburg-Landau theory. It is found that the enhanced resistivity is strongly influenced by the $c$-axis component of the magnetic field. This may be induced by the paramagnetically limited superconductivity, or the inhomogeneity in the films. Our observations provide fruitful information for this newly discovered infinite-layer nickelate superconducting system.

Supporting note added: When this work is under review, a new manuscript is posted on arXiv. It reports similar transport measurements on the $Nd_{0.775}Sr_{0.225}NiO_2$ thin film. The authors reached the same conclusion as ours, namely the upper critical field is determined by the paramagnetic limiting effect [67].

**Conflict of interest**

The authors declare that they have no conflict of interest.

## Acknowledgements

This work is supported by the National Key R&D Program of China (Grant No. 2016YFA0300401 and No. 2018YFA0704202), the National Natural Science Foundation of China (Grant No. 12061131001, No. 11774153, and No. 1861161004), the Strategic Priority Research Program of Chinese Academy of Sciences (Grant No. XDB25000000), and the Fundamental Research Funds for the Central Universities (Grant No. 0213-14380167).

## Author contributions

Yueying Li and Yuefeng Nie grew the $Nd_{1-x}Sr_xNiO_3$ thin films. Qing Li and Yueying Li treated the films by the soft-chemistry topotactic reduction method. Ying Xiang and Huan Yang carried out resistivity measurements with assistance from Hai-Hu Wen. Huan Yang, Hai-Hu Wen and Ying Xiang analyzed the data and wrote the manuscript which was reviewed by all the authors.

# Figures

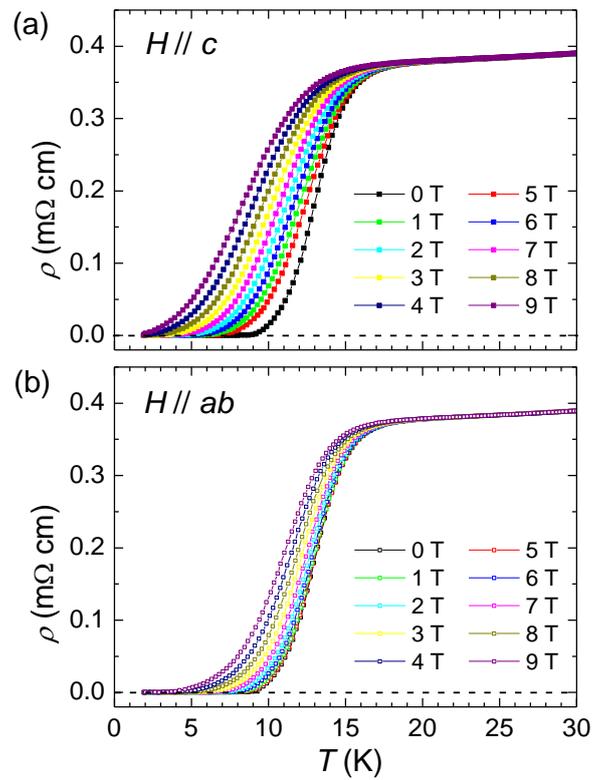

**Fig. 1.** (Color online) Temperature dependence of in-plane resistivity under different magnetic fields. Temperature dependent resistivity (the current $I \parallel ab$-plane) measured in the Nd$_{0.8}$Sr$_{0.2}$NiO$_2$ thin film at different magnetic fields ($H \perp I$) with (a) $H \parallel c$-axis and (b) $H \parallel ab$-plane, respectively.

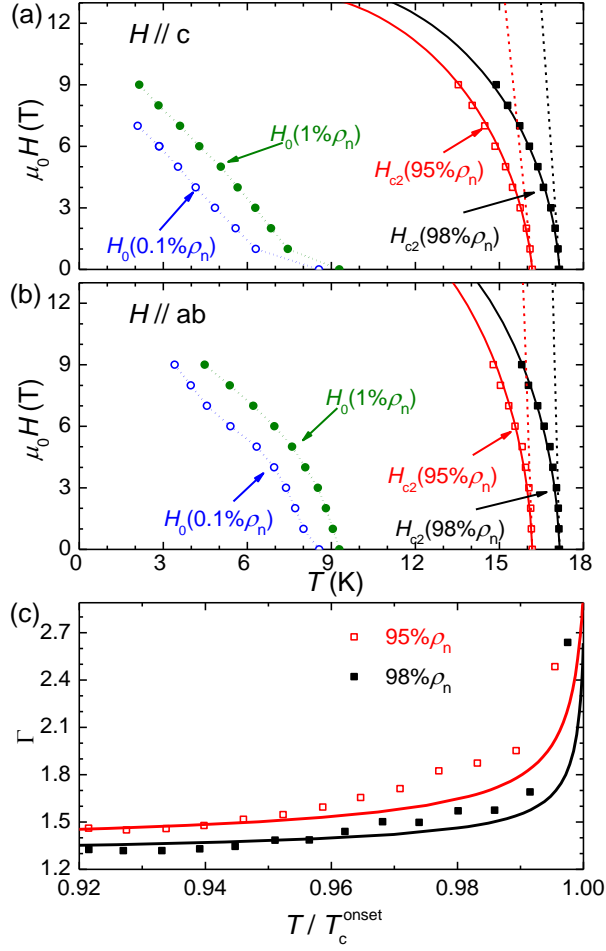

**Fig. 2.** (Color online) Superconducting phase diagram and superconducting anisotropy. Symbols in (a,b) show temperature dependent $\mu_0 H_0$ and $\mu_0 H_{c2}$ obtained from $\rho$-$T$ curves measured at different fields, and the sold (dashed) lines show the fitting results (theoretical curves) obtained from the WHH theory with finite (zero) values of $\alpha_M$. $\mu_0 H_0$ values are determined by using criterions of $0.1\%\rho_n(T)$ and $1\%\rho_n(T)$, while $\mu_0 H_{c2}$ values are determined by using criterions of $95\%\rho_n(T)$ and $98\%\rho_n(T)$. A clear negative curvature can be observed on the $\mu_0 H_{c2}(T)$ curves near the onset transition temperature. The solid lines are the fitting results to $\mu_0 H_{c2}$-$T$ data by using the WHH theory (Eq. 1) and assuming the absence of spin-orbital interaction ($\lambda_{SO} = 0$); the values of $\alpha_M$ are listed in Table 1. The dashed lines are the curves of the WHH theory with $\lambda_{SO} = 0$ and $\alpha_M = 0$. (c) Temperature dependence of the anisotropy ratio $\Gamma(T) = H_{c2}^{H||ab}(T)/H_{c2}^{H||c}(T)$ (symbols) calculated based on the $\mu_0 H_{c2}$ data in (a) and (b), and solid lines present $\Gamma$ values derived from the fitting curves in (a) and (b).

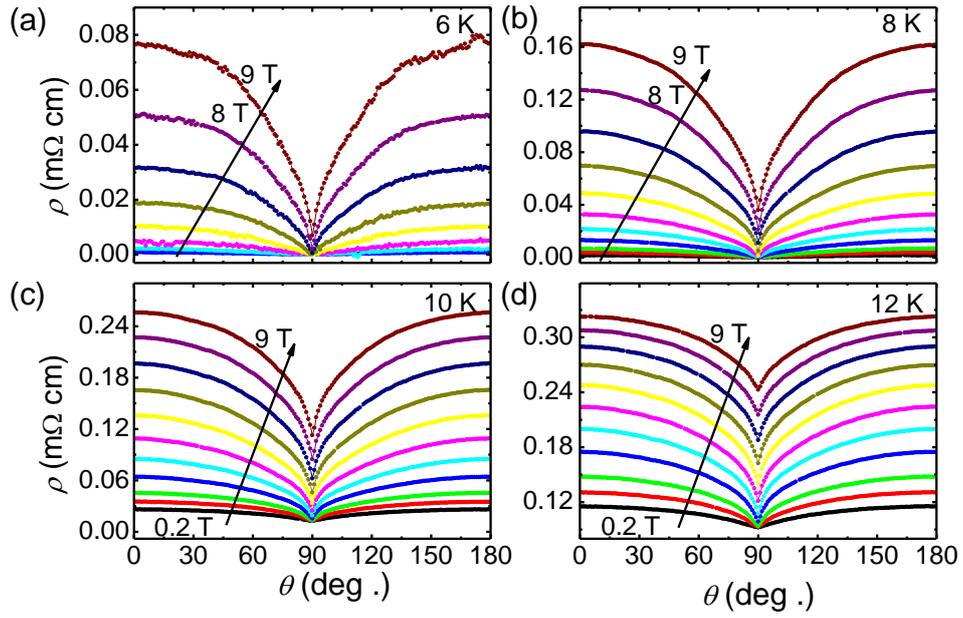

**Fig. 3.** (Color online) Angular dependent resistivity measured at different magnetic fields and temperatures. The direction of $\theta = 0°$ or $180°$ corresponds to the direction of $H \parallel c$-axis, while $\theta = 90°$ corresponds to that of $H \parallel ab$-plane. Magnetic fields are (a) from 2 to 9 T with an increment of 1 T, (b-d) 0.2, 0.5 T and from 1 to 9 T with an increment of 1 T.

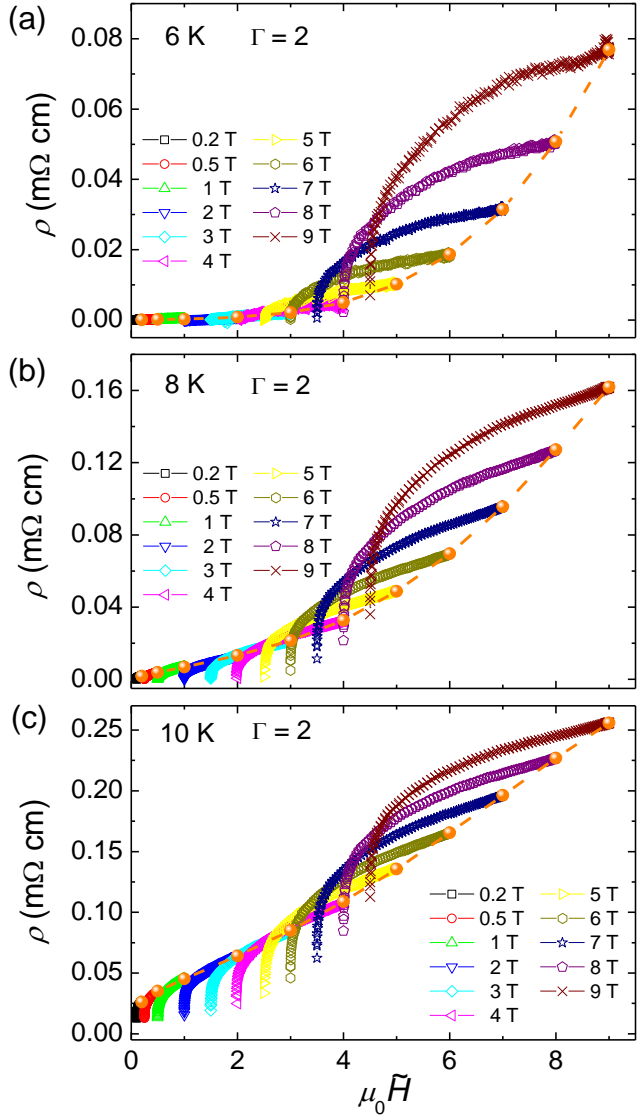

**Fig. 4.** (Color online) Scaling results based on the anisotropic GL theory. Examples of the failed scaling results to the $\rho$-$\theta$ curves by using the anisotropic GL theory. Here $\widetilde{H} = H\sqrt{\cos^2\theta + \Gamma^{-2}\sin^2\theta}$. Dashed lines connecting orange sphere-shaped symbols show the resistivity measured at $\theta = 0$ degree, and the scaling curves should coincide to these lines if the anisotropic GL theory worked.

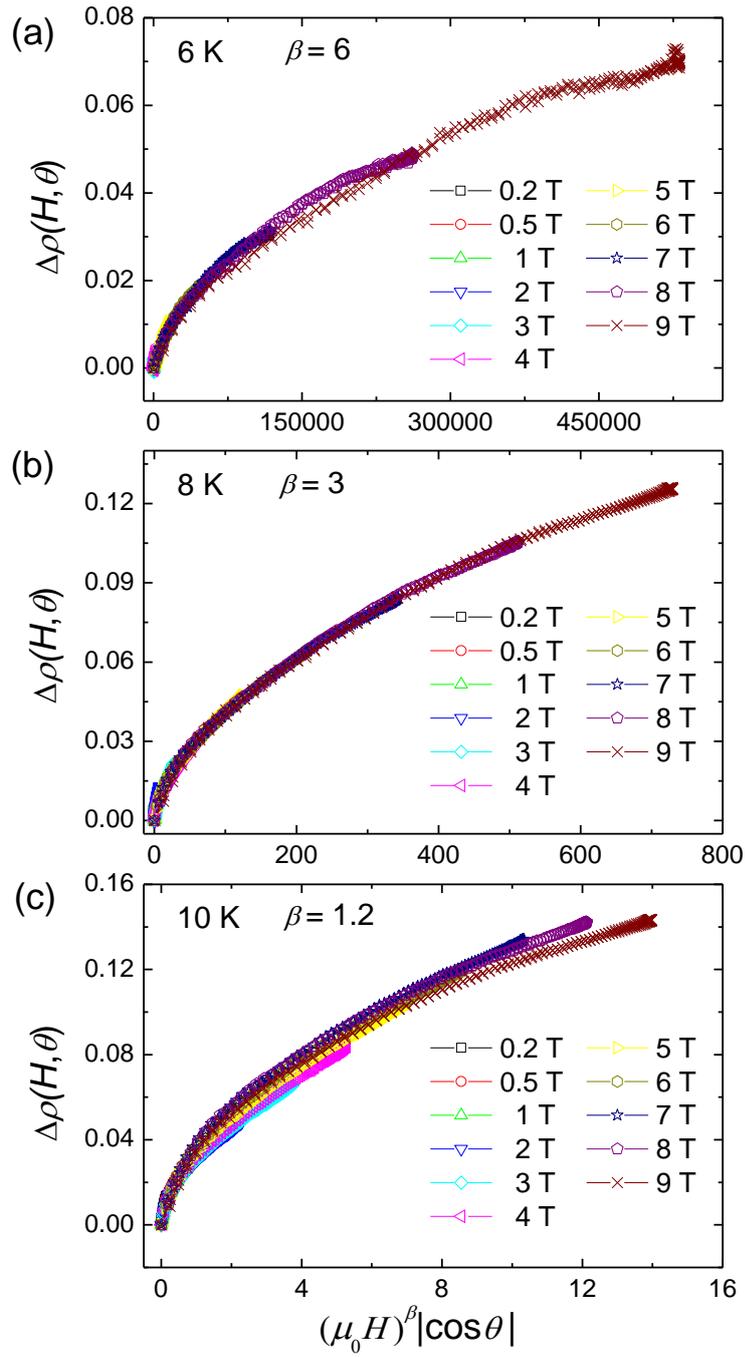

**Fig. 5.** (Color online) Scaling results based on a new scaling method. Scaling results of the resistivity difference $\Delta\rho(H, \theta) = \rho(H, \theta) - \rho(H, \theta = \pi/2)$ versus $(\mu_0 H)^\beta |\cos\theta|$ at different temperatures.

**Tables**

| Criterion | $T_c^{onset}$ | Field direction | $H'$ | $\mu_0H_{c2}^{orb}(0)$ | $\Gamma_{orb}(0)$ | $\alpha_M$ | $\mu_0H_P(0)$ | $\Gamma_P(0)$ |
|---|---|---|---|---|---|---|---|---|
| 98%$\rho_n$ | 17.1 K | $H \parallel c$ | 19.9±0.7 T/K | 235±9 T | 2.63±0.20 | 20±1 | 16.6±0.8 T | 1.25±0.09 |
| | | $H \parallel ab$ | 52.4±3.5 T/K | 619±42 T | | 42±2 | 20.8±1.0 T | |
| 95%$\rho_n$ | 16.2 K | $H \parallel c$ | 13.4±0.3 T/K | 149±4 T | 2.97±0.16 | 13±0.5 | 16.2±0.6 T | 1.29±0.08 |
| | | $H \parallel ab$ | 39.7±2.0 T/K | 443±23 T | | 30±1.5 | 20.9±1.0 T | |

**Table 1.** $\mu_0H_{c2}(0)$ and $\Gamma$ calculated based on the WHH theory. The fitting is operated in the absence of spin-orbit scattering. Error bars of $H'$ are determined from the temperature uncertainty and the error in the process of acquirement of $H_{c2}(T)$; the error bars of $\alpha_M$ are determined by the parameter ranges with which theoretical curves can fit the experimental data well; the error bars of $\mu_0H_{c2}^{orb}(0)$, $\mu_0H_P(0)$ and $\Gamma_P(0)$ are determined from the error transfer formula.